\documentclass[manuscript]{aastex}     

\slugcomment{Submitted to AJ}

\shorttitle{The Erupting Dwarf Nova in M15}
\shortauthors{Shara et al.}

\begin{document}

\title{The Luminous Erupting Dwarf Nova CV1 in the Dense Globular Cluster M15\footnotemark[1] $^,$\footnotemark[2]}

\footnotetext[1]{Based on observations with the NASA/ESA {\it Hubble Space Telescope},
obtained from the Data Archive at the Space Telescope Science Institute,which is 
operated by the Association of Universities for Research in Astronomy, Inc., 
under NASA contract NAS5-26555.}

\footnotetext[2]{We acknowledge with thanks the variable star observations from the AAVSO
International Database contributed by observers worldwide and used in this research.}

\author{Michael M. Shara}
\affil{Department of Astrophysics, American Museum of Natural History, 79th St. and Central Park West, New 
York, NY, 10024}

\email{mshara@amnh.org}

\author{Sasha Hinkley}
\affil{Department of Astrophysics, American Museum of Natural History, 79th St. and Central Park West, New
York, NY, 10024}
\email{shinkley@amnh.org}

\author{David R. Zurek}
\affil{Department of Astrophysics, American Museum of Natural History, 79th St. and Central Park West, New
York, NY, 10024}
\email{dzurek@amnh.org }

\author{Christian Knigge}
\affil{Department of Physics and Astronomy, University of Southampton,Southampton SO17 1BJ, United Kingdom} 
\email{christian@astro.soton.ac.uk}


\date{Received / Accepted}

\begin{abstract}
Despite decades-old predictions of the expected presence of dozens of 
cataclysmic variables in the cores of globular clusters, the number of
irrefutable, out-bursting candidates is still barely a handful.
Using multi-wavelength, multi-epoch HST images we have produced outburst 
and quiescence light curves for the recently discovered large amplitude 
variable CV1 in the core of the post core-collapse globular cluster M15. 
The light curves and blue colors show that the object is a bona fide 
dwarf nova, with absolute magnitude at maximum light rivaling that of 
the most luminous known dwarf novae.
\end{abstract}




\section{Introduction}

The realization that X-ray binaries are 100-1000 times overabundant in 
Galactic globular clusters relative to the rest of the Milky Way 
\citep{cla75,kat75} was crucial in understanding that intimate and tightly 
coupled relationships exist between stellar evolution and stellar 
dynamics. Among the predictions of this theory is the expectation that 
many white dwarf/red dwarf binaries should form via two-body tidal 
capture \citep{fab75} and/or three-body exchange captures \citep{hut83,hut84}. 
Some of these binary systems will be in very close orbits, such that the red 
dwarf fills its Roche lobe. This leads to the transfer of mass, via an accretion disk, 
from the red dwarf to the white dwarf component -- in other words, to 
cataclysmic variables (CVs). Yet searches for the many dozens of expected
erupting CVs have typically found zero to two dwarf novae in clusters 
searched from the ground \citep{sha94,tua03} and with the Hubble Space Telescope 
(HST) \citep{sha96,kni02}. 

Recently, Chandra X-ray Observatory and HST observers claimed the detections of 
dozens of cataclysmic variable (CV) candidates in the cores of several
globular clusters. These candidates are X-ray bright in Chandra observations 
\citep{gri01} or UV-bright in HST observations \citep{kni02} 
and clustered in the cores of their host clusters. However, only zero or one  
of these UV or X-ray sources per cluster has yet exhibited the degree of variability  
(2 magnitude or larger outbursts) that originally led to labeling 
these variables as ``cataclysmic.''

About half of all the CVs known
in the solar neighborhood exhibit large amplitude outbursts --- the Dwarf 
Novae. In the absence of spectroscopic confirmation (very challenging to 
obtain) and/or large amplitude outbursts, the X-ray and UV-bright CV candidates in 
globular clusters may remain just that: candidates, for the foreseeable future. 
It is thus important to fully substantiate every claim of large amplitude 
variability in globular cluster stars. Only a few irrefutable, spectrographically 
confirmed CVs (e.g. \citet{gri95,edm99}), and cataclysmics confirmed by slitless far-UV spectroscopy 
\citep{kni03,kni04}, (including 2 good candidates in \citet{edm03}) are known in globular 
clusters. A recent addition to this exclusive club of erupting CVs is the variable 
labelled V1 in the dense globular cluster M15 (Charles et al. 2002, hereafter C02). 

M15 is one of the densest globular clusters in the Galaxy. It has probably undergone 
core collapse and almost certainly contains a significant amount of non-luminous 
matter in its core \citep{bau03}. The cluster hosts the remarkable low mass X-ray binary
(LMXB) 4U2127+119, whose optical 
counterpart \citep{aur84} is the eclipsing binary AC211 \citep{ilo87}. 
M15 has been extensively imaged by HST, with which observers have discovered 
a sharp central density cusp \citep{guh96}, very blue stars and mass 
segregation in the core \citep{dem96}, Blue Stragglers \citep{yan94} 
and optically identified millisecond pulsars \citep{and90}---all indicators of strong dynamical 
interactions. If confirmed as an erupting CV, the candidate dwarf nova  
(C02) would be one of the rarest of exotic globular cluster stars, and an 
object worthy of further, intensive study.
In anticipation fo our key result --- that V1 really is an erupting dwarf nova -- we will
hereafter refer to it as CV1. This is to avoid confusion with the nomenclature used in the 
Sawyer-Hogg (1973) catalog of variables in globular clusters, recently updated by \citet{cle01}. 
There is an entirely different variable star labeled V1 (in M15) in these catalogs. 

\section{The M15 Dataset}

M15 has been imaged with the HST Wide Field Planetary Camera (WFPC2) during 6 epochs, 
during one epoch with the HST Faint Object Camera (FOC), two epochs using the HST 
Space Telescope Imaging Spectrograph (STIS), and finally with the HST Advanced Camera for 
Surveys (ACS) for two epochs. The wide range of filters and long time base of the observations (a 
decade) are particularly useful in 
characterizing variable stars. We downloaded all available images of M15 
from the HST archive, and summarize the dataset in Table 1 and Figure 1. The variable is
seen in eruption in the 4th and 5th epochs (see Figure 2), and in quiescence (below our frame limits) 
in the 1st, 2nd, 3rd, 6th, 7th, 8th, 9th, and the 11th epochs. It is detected in 
the near UV in the 10th epoch, possibly in transition between eruption and quiescence. 
There is Planetary Camera (0.044"/ pixel)coverage of CV1 in the 1st, 2nd, 4th and ninth 
epochs. The 5th and 6th epochs have only WF2 (hence lower resolution, 0.1"/pixel) 
coverage. The 10th and 11th epochs were taken with ACS for our own 
survey for UV variables. Thumbnail images of the vicinity of CV1 in most 
of the filters used in each epoch are presented in Figure 1; the 
variable is indicated in each frame.

\subsection{Photometry of the Dwarf Nova Candidate}

All of the WFPC2, ACS, and FOC data were obtained from the STScI Data Archive while the STIS images 
were still in the proprietary stage. All images were subject to the normal STScI on-the-fly 
recalibration pipeline software. The individual WFPC2 images were combined using the ``Montage2'' 
routine contained within the stand-alone DAOphot package. These frames were run through DAOphot's 
matching program ``DAOmaster'' to derive the subpixel frame-to-frame shifts and thereby register all 
images to the first image of the first epoch. These shifts were passed to Montage2, which produced a 
single, sky subtracted, high signal-to-noise image for each epoch and filter. A variability search 
was performed by blinking rapidly through groups of four of the eleven epochs to look for any subtle 
changes in the appearance of the stellar field. Dwarf novae should rise from invisibility to 
easily detectable in the timeframe defined by the observations. Brightening from 
previously empty regions of the sky is easily seen. Aside from CV1 and the three other object 
mentioned in C02, no other DN-like behavior was observed.  

Accurate measurement of the dwarf nova candidate's brightness is 
nontrivial because it is crowded between two neighboring stars. We have 
found that the most accurate way to carry out photometry on CV1 is to use an 
aperture large enough to encompass CV1 and its two companions. We measured the summed 
brightness of the two companions in epoch 1 (when CV1 is invisible), and 
verified in subsequent epochs (with CV1 in quiescence) that these two 
stars do not vary. We subtracted off the brightness of these two 
companions in subsequent epochs. All of the results presented in this paper  
utilize aperture photometry while accounting for the varying 
platescales between instruments.

The long-term light curve of CV1 is shown in Figure 3. We have assumed a distance 
modulus to M15 of 15.0 \citep{mcn04} and thereby transformed to the absolute 
magnitudes displayed in the Figure. The observations which use the ACS, FOC, and 
STIS instruments (with the F150LP, F165LP, F220W, F430W, and F28X50LP filters) are 
calibrated to the ST magnitude system, while the WFPC2 magnitudes are in the Vega system. 
Vega zeropoints are only available for the WFPC2 filters, and so the ACS, FOC and STIS 
photometry was left in the ST system. Reddening corrections have been imposed on the 
photometry using a reddening value for M15 of $E(B-V)=0.09$ \citep{cox00}. Using this value,  
the reddening corrections in the Figs. 3 and 5 were $A_{220} = 0.92$, $A_U = 0.41$, and 
$A_B = 0.36$ for the F220W, F336W, and F439W filters, respectively. Similar reddening corrections, 
appropriate for each band, were applied to the derived platelimits in Fig. 3. 
CV1 was fainter than the WFPC2 frame limit corresponding to absolute 
magnitude $\sim 6.0$ in each of the April 1994 and August 1994 images. The variable 
reached maximum brightness as seen in the F336W filter on 26 October 
1994, at absolute magnitude +1.5. It was next imaged in December 1998 
for seven consecutive days in multiple filters, when it was also in 
eruption. We have assembled the photometry for every available image during 
this week in Figure 4. There are indications that the variable was dimming towards
the end of this epoch. CV1 was below our detection threshold in the remaining epochs, 
except for a detection in the near UV with the F220W filter on ACS on 27 October 2003.

\section{The Eruptions of CV1}

Charles et al (2002) discovered CV1 in eruption in U band (F336W) HST PC 
images of October 1994 (seen in Figure 1 as epoch 4). They predicted 
that the object is blue, and suggested (since X-ray transients erupt 
only rarely) that CV1 is much more likely to be a dwarf nova than an 
X-ray transient.

The December 1998 images (epoch 5) all have CV1 on the (lower resolution) HST WF. 
CV1 is somewhat blended with its nearest neighbors, but careful measurement 
of the position of CV1 (see also Figure 2) shows that it is resolved and seen in 
eruption in F336W during the 1998 observations (contrary to the claim by C02). 
Comparison of the F336W and F439W images of CV1 in this epoch (Figure 1) clearly 
show that CV1 is very blue, with $U-B \sim -0.53 $ as predicted by C02. The light curve 
(and peak eruption luminosity) of CV1 shown in Figures 3 and 4 are very similar to the 
typical eruption light curve and luminosity of the prototypical dwarf nova SS Cygni centered on 
Julian Date 24501226, shown in Figure 5. The $U-B$ color is also similar to that 
of SS Cyg \citep{hop84} in outburst. The apparently significant flickering seen in the 
F502N and the H$\alpha$ images (Figure 4) is often seen in erupting dwarf novae. 

SS Cygni erupts, on average, every 40 days, and is brighter than minimum light for 
16 days during this time \citep{szk84}. It is nearly at maximum for 7 days, 
followed by a 7 day long decline to minimum. If CV1 is like SS Cygni then we should 
see it above its minimum state $\sim 16/40 = 40\%$ of the time. In fact we detect 
CV1 in eruption or above minimum in 3 of 11 epochs, or 27\% of the time --- an entirely
consistent result, given the small number statistics. 

Finally, CV1 is faint but visible in the F220W images of epoch 10, at 
absolute magnitude 5.7, about 4.0 magnitudes fainter than in epochs 4 
and 5 (ignoring color effects). If we assume a UV color (F220W -- $U) \sim -0.5$ then
CV1 is still 3.5 magnitudes fainter in epoch 10 than in epochs 4 and 5.
We suggest that this is an intermediate state between the star's quiescent 
and outburst states. This is because CV1 is NOT seen in the far-UV images 
(of epoch 11), which should be even more sensitive to UV-bright CVs than the 
near-UV imagery of epoch 10.

\subsection{Is CV1 a Dwarf Nova?}

The absolute magnitude of CV1 at maximum light, coupled with its very 
blue color, similarity in light curve to SS Cygni near maximum, 
detection in near UV but not in visible light near minimum, relative 
frequency of eruption (at least twice and possibly three times in eleven 
independent epochs) and amplitude of eruption (at least 4 and possibly 
more magnitudes) are all strongly supportive of CV1 being a dwarf nova.

While the case is not perfect (we do not have confirmatory spectra or a 
full light curve from quiescence to outburst and back again) it is very 
strong, and we fully concur with the characterization of C02 of CV1 as a dwarf nova.

We predict that an HST outburst spectrum of CV1 would show Balmer 
absorption lines, but both the crowding of CV1 in particular and the 
unpredictability of dwarf nova outbursts in general will make this a 
challenging observation (the $S/N$ for the the $H\alpha$ imagery of 
epoch 5 is too low to test this prediction).

While M15 now appears to possess a dwarf nova, it remains the only one known
in this dense, massive cluster. The two week duration search for dwarf novae in 
M15 by \citet{tua03} failed to detect any new candidates. The remarkable result that 
either zero, one or at most two erupting dwarf nova can be found in massive globular 
clusters seems to be holding true.


\acknowledgments


\appendix 


\clearpage



\begin{figure}
\figurenum{1}
\plotone{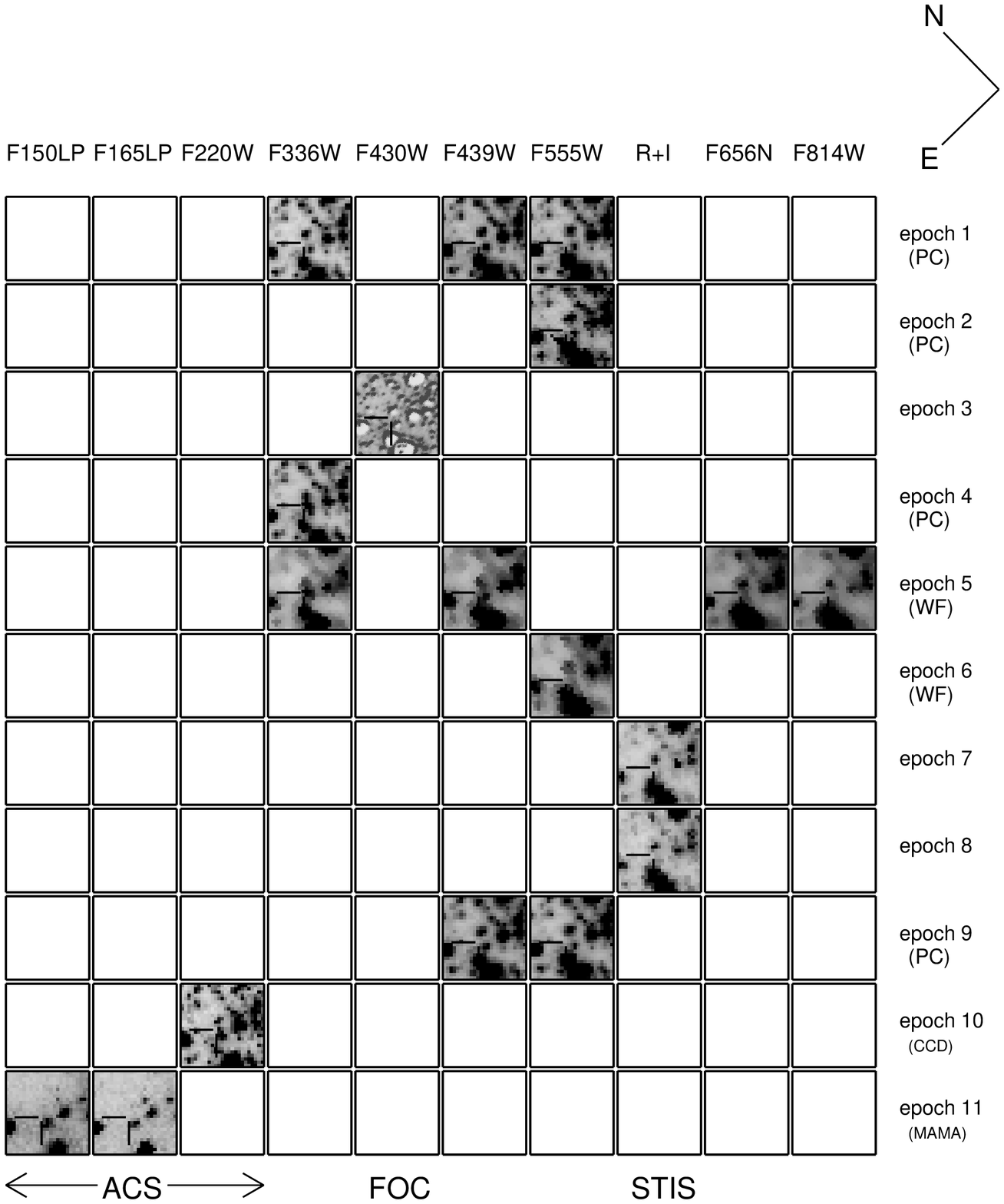}
\caption{A matrix of HST images of the field of the candidate dwarf nova CV1 in M15. North and East 
are as indicated, and each postage stamp image is 1.24 arcsec on a side. The upper labels are 
the HST filter names. CV1 is indicated in each postage stamp image.}
\end{figure}

\begin{figure}
\figurenum{2}
\plotone{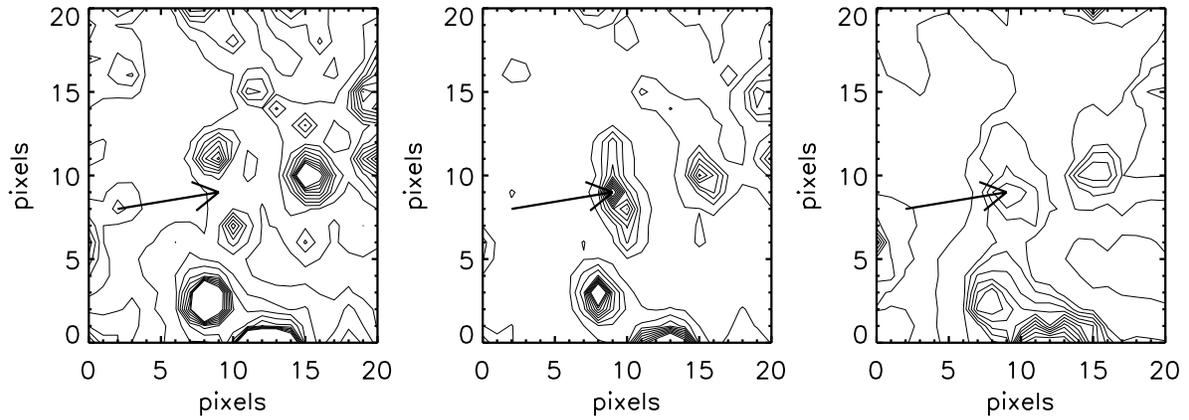}
\caption{A contour plot showing the neighborhood of the erupting Dwarf Nova CV1. From left to right,
the plots show the first (quiescent), and fourth and fifth (eruptive) epochs, respectively in F336W. The arrow shows the location of the
Dwarf Nova. Each plot is 1.09 arcsec on a side. The contours of the left and middle plots indicate 40-count intervals, while 
the right plot shows 20 counts per contour. }
\end{figure}  

\begin{figure}
\figurenum{3}
\plotone{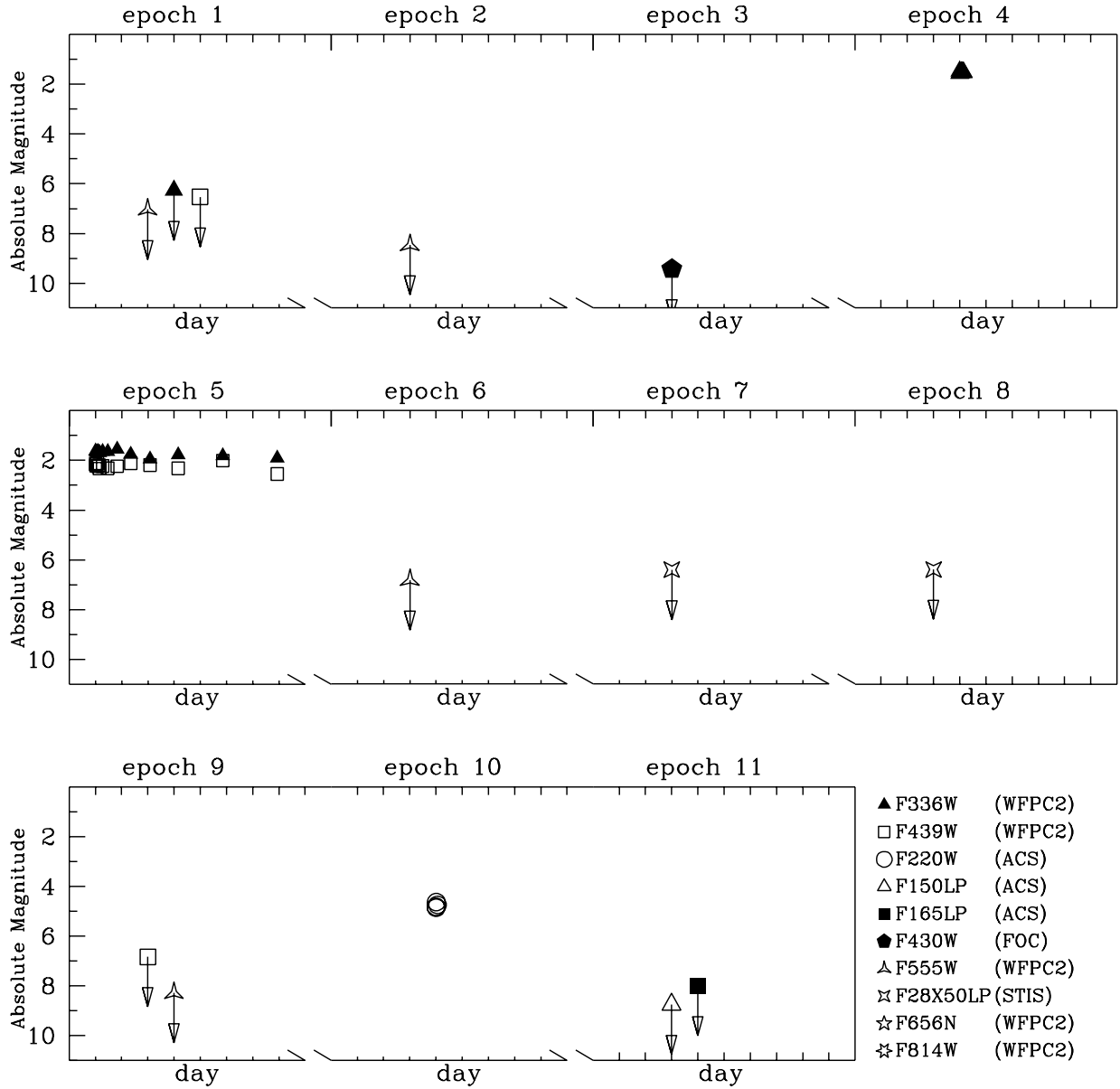}
\caption{Eleven epochs of multi-wavelength photometry of CV1 in M15.  The object is in eruption in epochs 4 and 5,
and probably in an intermediate state in epoch 10. Also shown are the upper--detection limits for those epochs in 
which CV1 is quiescent. }
\end{figure}

\begin{figure}
\figurenum{4}
\plotone{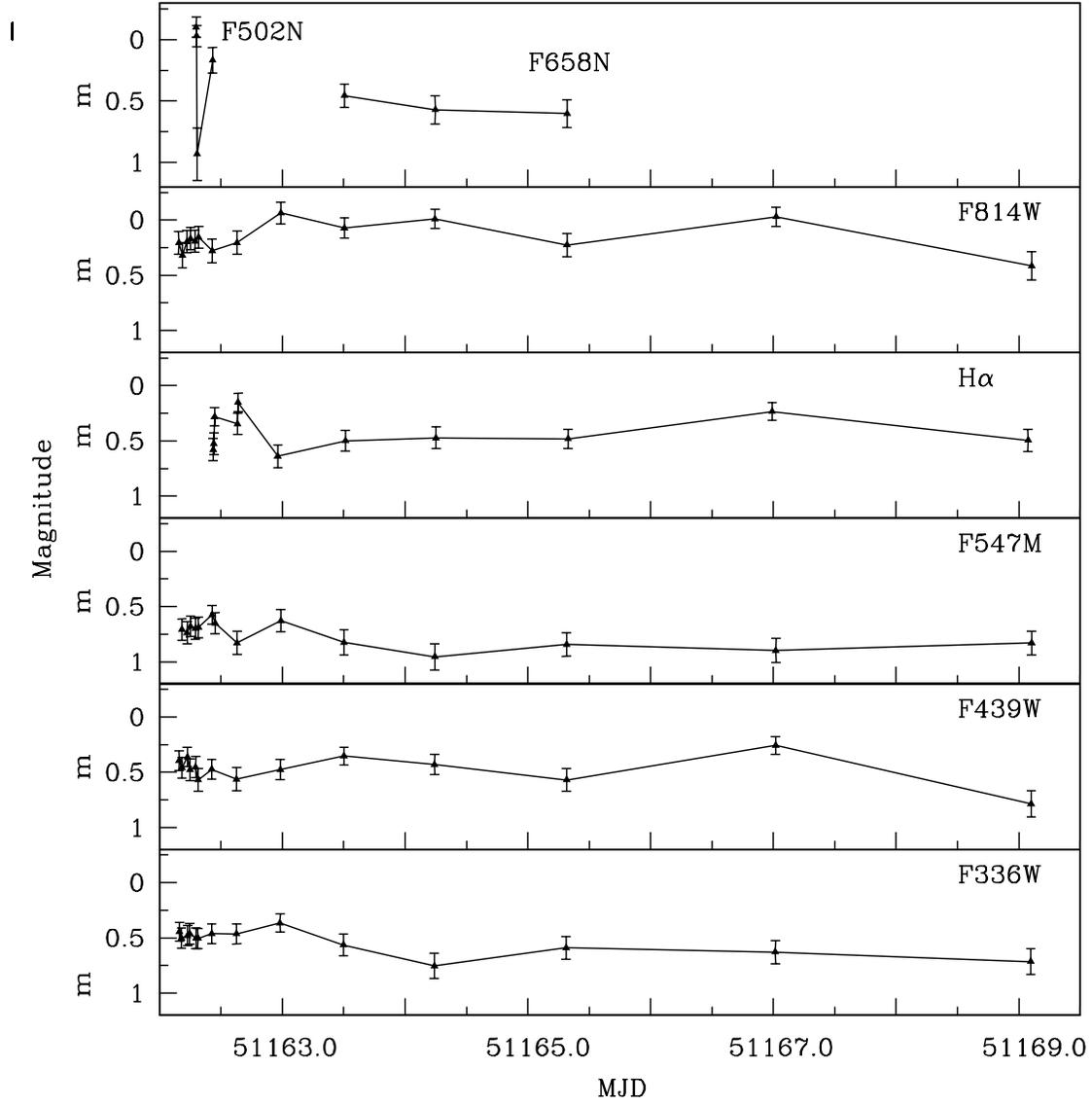}
\caption{Daily observations of the dwarf nova CV1 in outburst over 1 week in December 1998 (epoch 5). Magnitude offsets are arbitrary.}
\end{figure}

\begin{figure}
\figurenum{5}
\plotone{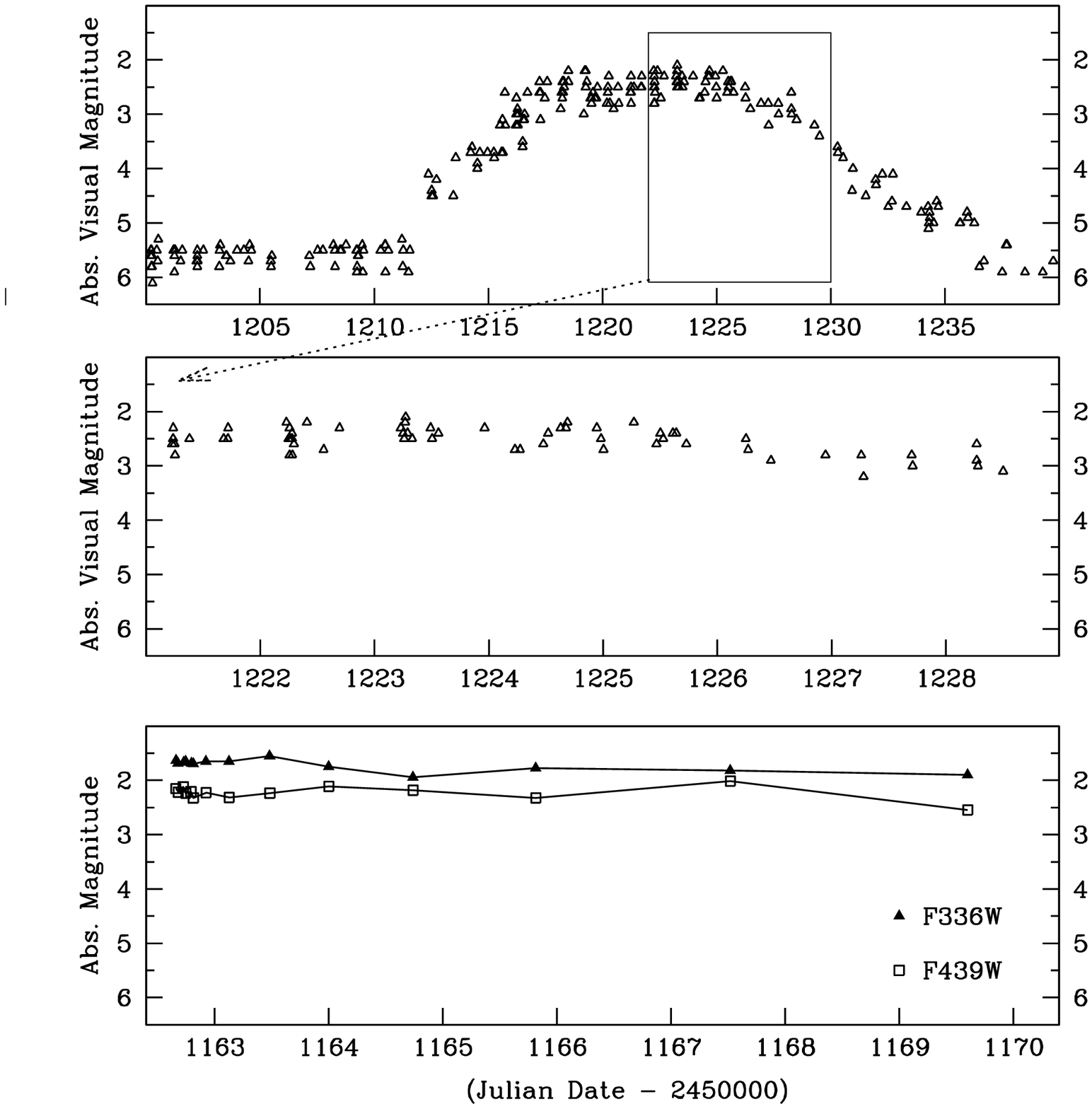}
\caption{ The dwarf nova SS Cygni in outburst over two different time resolutions (upper two plots). 
Observations are visual, from the AAVSO. Also shown 
is the lightcurve of  CV1 in M15 (same as Figure 4) in F336W and F439W. The light curves suggest that CV1 
was in early decline from a dwarf nova outburst in epoch 5. }
\end{figure} 

\begin{deluxetable}{l|l|c|c|c|c|c|c|c|c|c|c}
\tabletypesize{\scriptsize}
\rotate
\tablecaption{Table of M15 observations: Listed are the number of useable observations at each date. The column labels from 
F150LP to F814W are HST filter names, in order of increasing central wavelength. Also, 
for the WFPC2 images, the placement of the center of the cluster falls either on the PC chip or a WF, as indicated.}
\tablewidth{0pt}
\tablehead{\colhead{Epoch} & \colhead{PI/Prog \#} & \colhead{F150LP} & \colhead{F165LP} & \colhead{F220W} & 
\colhead{F336W} & \colhead{F430W} & \colhead{F439W} & \colhead{F555W} & \colhead{R+I} & \colhead{F656N} & \colhead{F814W}}
\startdata
               &                  &        ACS        &         ACS       &       ACS         &        WFPC2          &       FOC         &       WFPC2         &        WFPC2         &     STIS       &          WFPC2       &     WFPC2           \\ \hline
1:  04/07/94   &Yanny/5324        &                   &                   &                   &2 PC$\rightarrow$200s  &                   &2 PC$\rightarrow$80s &4 PC$\rightarrow$32s  &                &                      &                     \\ \hline
2:  08/30/94   &Bahcall/5687      &                   &                   &                   &                       &                   &                     &2 PC$\rightarrow$400s &                &                      &                     \\ \hline 
3:  09/27/94   &King/5301         &                   &                   &                   &                       &1$\rightarrow$2021s&                     &                      &                &                      &                     \\ \hline 
4:  10/26/94   &Westphal/5742     &                   &                   &                   &11 PC$\rightarrow$5000s&                   &                     &                      &                &                      &                     \\ \hline 
5a: 12/15/98   &Bond/6751         &                   &                   &                   &9 WF$\rightarrow$207s  &                   &9 WF$\rightarrow$180s&                      &                &6 WF$\rightarrow$3820s&9 WF$\rightarrow$126s\\  
5b: 12/16/98   &     ""           &                   &                   &                   &1 WF$\rightarrow$23s   &                   &1 WF$\rightarrow$20s &                      &                &1 WF$\rightarrow$1400s&1 WF$\rightarrow$14s \\  
5c: 12/17/98   &     ""           &                   &                   &                   &1 WF$\rightarrow$23s   &                   &1 WF$\rightarrow$20s &                      &                &1 WF$\rightarrow$1400s&1 WF$\rightarrow$14s \\  
5d: 12/18/98   &     ""           &                   &                   &                   &1 WF$\rightarrow$23s   &                   &1 WF$\rightarrow$20s &                      &                &1 WF$\rightarrow$1400s&1 WF$\rightarrow$14s \\  
5e: 12/19/98   &     ""           &                   &                   &                   &                       &                   &                     &                      &                &1 WF$\rightarrow$1700s&                     \\ 
5f: 12/20/98   &     ""           &                   &                   &                   &1 WF$\rightarrow$23s   &                   &1 WF$\rightarrow$20s &                      &                &                      &1 WF$\rightarrow$14s \\ 
5g: 12/22/98   &     ""           &                   &                   &                   &1 WF$\rightarrow$23s   &                   &1 WF$\rightarrow$20s &                      &                &1 WF$\rightarrow$1700s&1 WF$\rightarrow$14s \\ \hline 
6:  08/31/99   &Van Altena/7469   &                   &                   &                   &                       &                   &                     &12 WF$\rightarrow$312s&                &                      &                     \\ \hline 
7:  10/21/99   &Van der Marel/8262&                   &                   &                   &                       &                   &                     &                      &4$\rightarrow$8s&                      &                     \\ \hline  
8:  10/22/01   &Van der Marel/8262&                   &                   &                   &                       &                   &                     &                      &2$\rightarrow$4s&                      &                     \\ \hline 
9:  04/05/02   &McNamara/9039     &                   &                   &                   &                       &                   &4 PC$\rightarrow$160s&12 PC$\rightarrow$192s&                &                      &                     \\ \hline 
10: 10/27/03   &Knigge/9792       &                   &                   &5$\rightarrow$2900s&                       &                   &                     &                      &                &                      &                     \\ \hline  
11: 11/13/03   &     ""           &3$\rightarrow$2410s&2$\rightarrow$2500s&                   &                       &                   &                     &                      &                &                      &                     \\ \hline 
\enddata
\end{deluxetable}

\end{document}